\documentstyle[prb,aps,preprint,epsfig]{revtex}

\begin{document}

\date{\today}
\draft


\title{
A theoretical analysis of Ballistic Electron Emission
Microscopy: k-space distributions and spectroscopy.
}

\author{
P.L. de Andres
}

\address{
Instituto de Ciencia de Materiales (CSIC),
Cantoblanco, E-28049 Madrid (SPAIN)
}

\author{
K. Reuter
}

\address{
Lehrstuhl fur Festk\"orperphysik,
University of Erlangen-N\"urnberg (Germany) 
}

\author{
F.J. Garcia-Vidal,
D. Sestovic and F. Flores
}

\address{
Dept. de Fisica Teorica de la Materia Condensada (UAM),
Universidad Autonoma de Madrid, E-28049 Madrid (SPAIN)
}

\maketitle

\begin{abstract}
We present a theoretical framework
well suited to analyze
Ballistic Electron Emission Microscopy (BEEM)
experiments.
At low temperatures and low voltages, near the threshold
value of the Schottky
barrier, the BEEM current
is dominated by the elastic component.
Using a Keldysh Green's functions method,
we analyze
the injected distribution of electrons
and
the subsequent
propagation through the metal.
Elastic scattering by the lattice
results in
the formation of focused beams
and narrow lines in real space.
To obtain the current injected in
the semiconductor,
we compute the current
distribution in reciprocal space
and,
assuming energy and $k_{\parallel}$
conservation,
we match states to the
projected conduction
band minima of the semiconductor.
Our results show an important focalization
of the injected electron beam and explain
the similarity between BEEM currents for
Au/Si(111) and Au/Si(100).
\end{abstract}

Sci., in press)\\

\section{INTRODUCTION}

Among the many exciting applications of the Scanning
Tunneling Microscope (STM) one has recently become
established as a technique in its own:
Ballistic Electron Emission Microscopy (BEEM)
\cite{kaiser,mario}. In the standard
version of this technique, the STM acts as a 
microscopic gun injecting a very narrow and coherent
beam of electrons on a metalic layer deposited
on a semiconductor. The electrons are subsequently
propagated through the metalic layer and finally
are detected in the semiconductor, passed the
metal-semiconductor interfacial Schottky barrier.
Both, spectral resolution and a lateral nanometric
geometrical resolution, allow a detailed study of
the buried interface that cannot be easily obtained
by other techniques. Besides the interest of
obtaining accurate information on the metal-semiconductor
interface, other processes of technological
importance have a considerable influence on the final
BEEM current, and can be studied by applying this
technique. A relevant example is the current attenuation
length, determined by various inelastic processes:
clearly these quantities are of paramount importance
to design and operate very small devices.

The main obstacle on this road is just the
complexity of the technique itself.
The general case is that many different
physical processes contribute in quite different
ways to the BEEM current, resulting in a
complex mixing of interfering effects that
preclude an easy determination of each of
them separately. The only way out is to
construct a realistic theory that can reliably
handle each factor, trying to avoid spurious correlations
of fitted values. 
Unfortunately, this is often the case when a
parametrized approach, like the popular E-space
Monte-Carlo technique, is applied to obtain
different values related between them,
transfering uncertainties and errors from
one place to the other without control.
In particular, the elastic interaction of
electrons with the lattice has been described as equivalent
to a random walk, that cannot
mimic the real interaction
that is known to produce gaps in certain
directions. Free propagation of
carriers on distances of
about $20$ \AA\ 
in forbidden directions are not 
desirable, but possible in
those Monte-Carlo simulations. 
Also, the simulations are very sensitive to
the initial tunneling distribution,
that is taken into account only through
planar tunneling theory and WKB approximation.

As mentioned before, an important capability
of the technique is its ability to measure
different attenuation lengths related to
inelastic interactions. The main inelastic
processes are the electron-electron and
the electron-phonon interaction. 
Lowering the temperature to $77$ K the
electron-phonon interaction is already
very low and one can ignore it to concentrate
on the electron-electron interaction only.
A very interesting result obtained using
this approach is that the standard lifetime
derived considering the dynamically screened
Coulomb potential in an electron gas with
a density related to gold cannot adequately
explain the BEEM experiments\cite{bell111}.
This result seems independent of the particular
set of parameters used in the E-space Monte-Carlo
simulation, and has been also confirmed by 
fully {\it ab initio} k-space Monte-Carlo\cite{ulrich}. 
Because of the importance of this problem,
we shall discuss it further in the context of
our {\it ab initio} Green's function calculation.

In this paper we describe an {\it ab initio} Green's
function calculation that can be used to compared
with experimental results.
In the pure elastic case
(ballistic current) this formalism does not use any
free parameter, while to take into account the
inelastic electron-electron interaction we make
use of a mean free path parametrization proposed
by Bell\cite{bell111}, that let us to choose a
single value to get good agreement with experiments.
We notice that this is a rather similar situation
to other fields, like the Low Energy Electron
Diffraction, where first principles Green's
functions methods are supplemented with
parametrized values of the imaginary part
of the self-energy (typically a constant
value) to include inelastic effects in the
theory\cite{pendry}. This approach has been
proven succesful in those fields, and we hope
it will introduce for BEEM a new way to compare 
theory and experiment.
The paper is organized as follows:
in section II we describe a theoretical
framework covering three important steps on a BEEM
experiment: the initial tunneling injection, 
the propagation of electrons through the metalic
layer, and the transmission of electrons at the
two-dimensional plane defining the interface
between the metal and the semiconductor.
Finally, in section III, our theoretical results
are compared with spectroscopic I(V) experimental
data taken
under conditions compatible with our main
assumptions by other authors.

\section{THEORY}

Our theoretical approach is based on a hamiltonian
describing separately the STM tip (T), the metal (M),
and their interaction (I):

\begin{equation}
\hat H = \hat H_{T} + \hat H_{M} + \hat H_{I}
\end{equation}

\noindent
where the tip is defined
by $\hat H_{T} = \sum \epsilon_{\alpha}
\hat n_{\alpha} + \sum \hat T_{\alpha \beta}
c_{\alpha}^{\dagger} c_{\beta}$,
the metal is given by
$\hat H_{M} = \sum \epsilon_{i}
\hat n_{i} + \sum \hat T_{i j}
c_{i}^{\dagger} c_{i}$, and
the coupling term between them is
$\hat H_{I} =
\sum \hat T_{\alpha j}
c_{\alpha}^{\dagger} c_{i}$
(where greek indexes, $\alpha, \beta$, describe
orbitals on the tip and latin
indexes, $i, j$, refer to the metal).
The interaction hamiltonian 
has been written 
in terms of a hopping matrix that
couples the atomic orbitals in the tip
with the atomic orbitals in the metal.
Because for BEEM work the tip-sample distance
is usually large, we make the approximation
of considering only tunneling between s-orbitals.
The hamiltonian describing the metal has
been written on a tight-binding approximation
with parameters given by Papaconstantopoulos\cite{constan}
for gold.

Currents between the tip and the
sample and within the metal base
are computed applying a Keldysh
formalism\cite{keldysh}:

\begin{equation}
J_{i j} =
\int d \omega Tr \{ \hat T_{i j}
(\hat G_{ij}^{+-} - \hat G_{ji}^{+-}) \} 
\end{equation}

\noindent
that is well suited for a non-equilibrium problem like
the tunneling one, but also allow to write an elegant
expresion for more standard scenarios like the
propagation through the metal layer. 
The central objects in this formalism,
$\hat G_{ij}^{+-}$, are obtained from a Dyson-like
equation through the retarded and
advanced Green functions of the
uncoupled parts of the system (the tip and the sample),
$\hat g^{R}$ and $\hat g^{A}$,
and the hopping matrix $\hat T$\cite{prl,phscr}. In that
context, it has been shown previously how the current
between two atoms $i, j$ in the metal can be obtained by:

\begin{equation}
J_{ij} = {2 e \over \pi \hbar} \Re \int_{-\infty}^{\infty}
d \omega \lbrack f_{T}(\omega) - f_{S}(\omega) \rbrack  \times
Tr \sum_{m \alpha \beta n} \lbrack
\hat T_{ij} \hat g_{jm}^{R} \hat T_{m \alpha}
( \hat g_{\alpha \beta}^{A} -
\hat g_{\alpha \beta}^{R} )
\hat T_{\beta n} \hat g_{ni}^{A}
\rbrack 
\end{equation}

\noindent
where $T_{ij}$ is the hopping matrix between atoms
$i,j$, $g_{jm}^{r}$ and $g_{mj}^{a}$ are the retarded
and advanced Green functions linking sites
$j$ and $m$, 
$( \hat g_{\alpha \beta}^{A} -
\hat g_{\alpha \beta}^{R} )$
is related to the
density of states matrix on the active atom at the
tip (from now on
assumed for simplicity to be $0$),
and the trace implies a summation over the
orbitals forming the chosen basis.
Eq. 3 describes the coherent propagation of electrons 
between the surface sites ($m$ or $n$), and
the atoms inside the crystal ($i$ and $j$),
and ingredient that we find of paramount
importance to describe appropriately the
BEEM current in the metal.

In this formalism the initial tunneling current is obtained
as the current between the last atom in the tip (0) and
all the atoms active for tunneling in the metal (m).
In the same Keldysh formalism, we can write:

\begin{equation}
J_{T}= \int d \omega Tr 
\sum_{m}[ \hat{T}_{0m} \hat{G}^{+,-}_{m0}(\omega) -
\hat{T}_{m0} \hat{G}^{+,-}_{0m}(\omega) ]
\end{equation}

\noindent
that for long tip-sample distances and
the simpler case where only one atom is
active for tunneling in the metal yields\cite{alvaro}:

\begin{equation}
J_{T} = \frac{4 \pi e}{\hbar}
\int_{-\infty}^{\infty} Tr [ \hat{T}_{01}
\hat{\rho}_{11}(\omega) \hat{T}_{10} \hat{\rho}_{00}(\omega)]
[f_{T}(\omega)-f_{S}(\omega)] d \omega
\end{equation} 

\noindent
where the further simplification of allowing
tunneling to only one atom in the metal
(m=1) has  been made. This formula
shows the dependence of the tunneling current
with the hopping matrix linking the tip to the
surface, $T_{01}$, the density of states at
the two active sites, $\rho_{00}$ and $\rho_{11}$,
and the different Fermi distribution functions.
This approach can be used to compute 
topographic images of a given surface
(either at constant intensity or
constant height) and already
has been succesfully used to compare
with experiments\cite{fernando}.

The current distribution in reciprocal space is necessary
to match wavefunctions across the 
metal-semiconductor interface in order to
calculate the BEEM current in the semiconductor.
It is possible to show that in k-space the current
per energy unit
between \underline{two layers $a, b$} in the metal,
is given
by an expression formally identical to 
Eq. 3.:

\begin{equation}
J_{ab}(k_{\parallel}) = 
{2 e \over \pi \hbar} \Re 
Tr \lbrack
\hat T_{ab} (k_{\parallel})
\hat g_{b1}^{R} (k_{\parallel}) \hat T_{10}
\hat \rho_{00}
\hat T_{01} \hat g_{1a}^{A} (k_{\parallel})
\rbrack 
\end{equation}

\noindent
where the various quantities are 
two-dimensional
Fourier transforms of the respective objects 
appearing in the real space current between 
\underline{two
nodes} of the metalic lattice $i, j$:

\begin{equation}
\hat T_{i,j} (\vec r_{\parallel}) =
\sum_{k_{\parallel}} 
e^{- i \vec k_{\parallel} \vec r_{\parallel}}
\hat T_{a,b}  (\vec k_{\parallel})
\end{equation}

\noindent
The summation is perfomed over a set of
special points covering the two-dimensional
Brillouin zone\cite{rafa}.

A proper calculation of the quantum mechanical transmission
coefficient, $T$, across the interface can be done by projecting
and matching 
metal and semiconductor
bulk states into the interface\cite{stiles}.
This procedure yields, as expected,
a square root variation
of that coefficient with the energy
(measured w.r.t. the semiconductor conduction band minimum).
Usually, a much easier model is applied in the literature
where the transmission coefficient over a step-like
potential barrier (taken as the Schottky barrier)
is considered: this brings the correct behaviour with
energy, but because this is a one-dimensional problem
the variation of $T$ with $k_{\parallel}$ is not well
represented. An intermediate level of sophistication,
improving significantly the model,
would be to describe the semiconductor within the Jones
zone
approximation and to match the wavefunctions 
logarithmic
derivatives given a particular orientation for the
semiconductor. In particular, this approach is quite
reasonable around the 
semiconductor energy gap and can describe satisfactorily
the region close to the conduction band minima.
We have followed this approach using a Surface Green's
function matching formalism\cite{elices,moliner} 
(equivalent to the mentioned wave function matching),
considering the neighbourghood of the projected 
$\Gamma-X$ direction ($\overline{M}$) in the
Si(111) and Si(100) surfaces: these are the regions 
where the conduction band minima of Si are projected.
In the parallel direction, where the kinematic restriction
of conserving $k_{\parallel}$ and having enough energy
to be injected in the semiconductor is applied, a
simple parabolic approximation with effective masses
taken from the literature
is used. This defines an ellipsoidal region
around the minimum of the semiconductor conduction
band
where the two-dimensional transmission
coefficient described above
changes continously from its value at the center
of the ellipse (the origin where the $\vec k_{\parallel}$
is measured in the semiconductor) to zero at the
boundaries of that region. 
This approach increases transmission at the
interace by values between $\frac{15}{100}$
and $\frac{40}{100}$, depending on the energy.

For the particular 
case of a (111) orientation and in the neighbourhood
of the $\overline{M}$ point we have\cite{elices}:

$$
\gamma = {(1-\frac{2}{3} h L)(1-\frac{5}{6} h L) \over
          (1+\frac{2}{3} h L)(1+\frac{6}{6} h L) } 
$$

\noindent
where $h={ 2 \pi \sqrt{3}\over a}$, and
${1 \over L} = \sqrt{2 (E- {\kappa^{2} \over 2}}$.

\noindent
defining,
$$
a = {e-{\kappa^{2} \over 2} - v +
\sqrt{(e-{\kappa^{2} \over 2})^{2} - v^{2}} \over v}
$$
\noindent
with $v \approx 2 eV$ to represent the silicon
gap.
Finally, the reflection amplitude is given by:

\begin{equation}
r = {  
{1+\gamma \over 1-\gamma} -
{a-1 \over a+1} 
\over 
{1+\gamma \over 1-\gamma} +
{a-1 \over a+1} 
}
\end{equation}

\section{RESULTS}

\subsection{k-space and r-space distribution.}

In this section, our formalism is applied to the case
of Au grown on Si(111) or Si(100) orientations. From
previous structural work performed with LEED,
Auger and STM\cite{structure}, we assume that
Au grows on (111) crystalline directions, except
for the first few layers near the interface that
may present some disorder.

\begin{figure}
\epsfig{figure=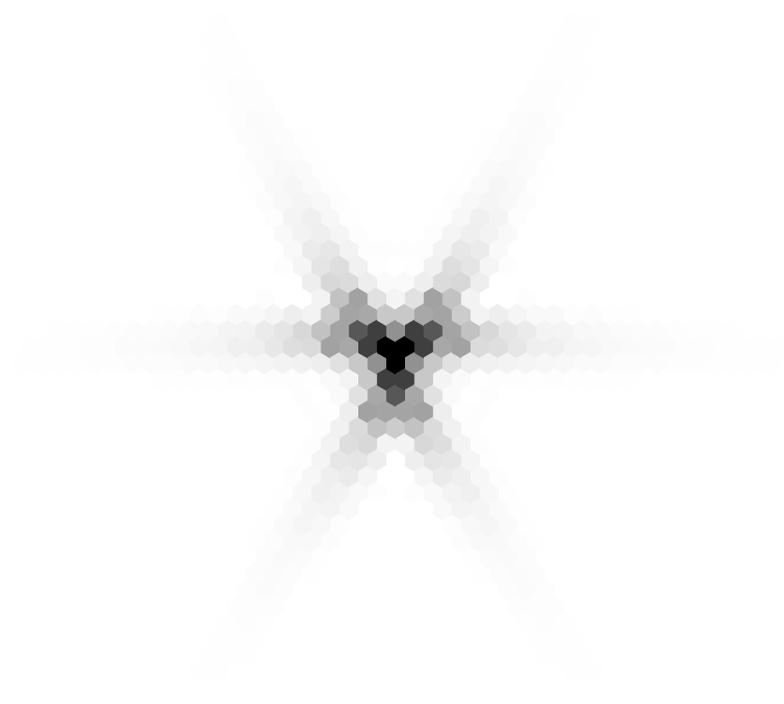,height=15cm,width=15cm}
\caption{
r-space distribution for Au(111).
5th layer (close to the surface),
}
\end{figure}

\begin{figure}
\epsfig{figure=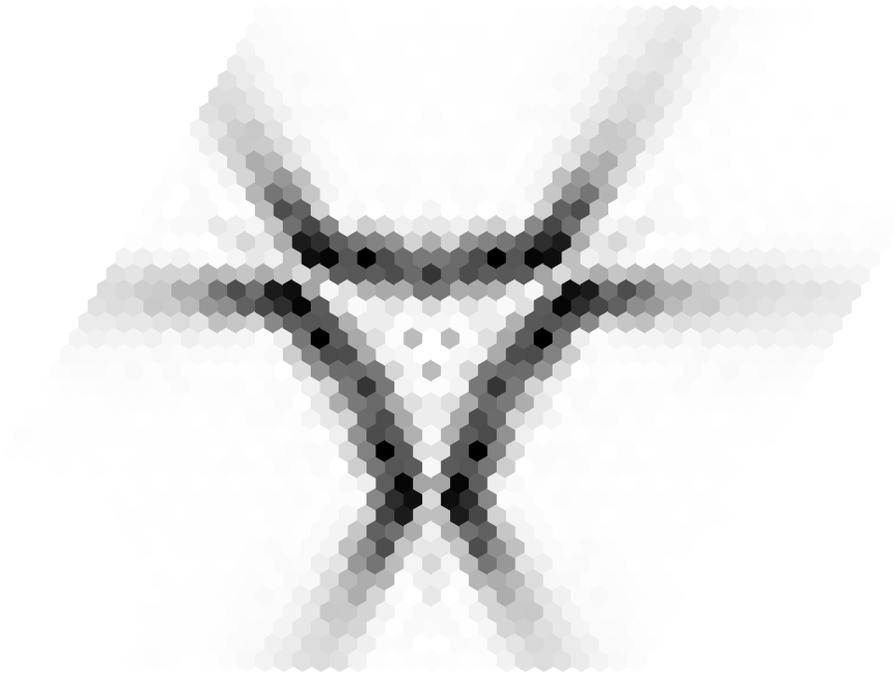,height=15cm,width=15cm}
\caption{
r-space distribution for Au(111).
15th layer (far from the surface).
}
\end{figure}

First of all, we observe the formation of narrow beams and
lines in real space (e.g. see Figs. 1 and 2). This effect was
previously predicted on the basis of a semi-classical
calculation of the Green's functions necessary to
compute Eq. 3\cite{prl}, and it is now confirmed working with a
more accurate, full quantum mechanical, approximation
based on a decimation technique\cite{guinea}.
The same effect has also been observed
when the propagation of electrons on (100) directions
is considered\cite{phscr}. In fact, within our
formalism, it is possible to follow the propagation
through the material layer-by-layer.
In this way, we can observe the gradual
construction of Bloch states inside the solid,
seen in particular by the formation of gaps in
characteristic directions like the (111) or (100)
depending on the energy. This is the case
when Figs. 1 and 2 are compared:
close to the surface (1) electrons
spread in all directions (including the (111)),
but far away (2) the propagation in that
direction is inhibited.

Secondly, we study the current distribution in k-space
(e.g. see Fig. 3). In the purely elastic case, the
energy is a real quantity and the wavefunction does not
decay. However, it is usual in any Green's function
formalism to add a very small imaginary part to the
energy to create a region in the complex plane free
of poles to ensure the proper analytic behaviour.
In this work, we discuss the case of a very small
imaginary part added to the energy ($\eta=0.001$ eV),
while we leave a detailed discussion of a finite $\eta$,
representing electron-electron
inelastic interaction, for a forthcoming
paper\cite{karsten}. A typical current distribution
under these conditions can be seen in Fig. 3.
This distribution does not present
noticeable changes when observed at different
layers.
The symmetry found is six-fold, also
obtained on
other similar quantum-mechanical
calculations\cite{stiles}. This result is
expected
because these currents reflect the projected density
of states on a given layer. 
It is interesting to notice, however, the
different symmetry found 
in real space (three-fold), 
reflecting the symmetry of the
whole lattice. 
 
\newpage

\begin{figure}
\epsfig{figure=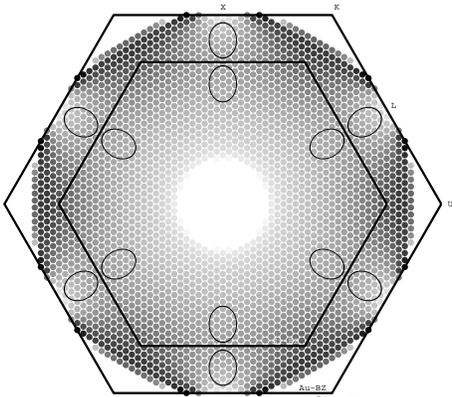,height=4cm,width=4cm}
\caption{
k-space distribution for Au(111), eta=0.001 eV.
}
\end{figure}

\subsection{Spectroscopy: Au on Si(111) and Si(100).}

An interesting feature of BEEM is its ability to
bring spectroscopic information of the interface.
A very simple model shows how the spectroscopy is
an integral on energies of different factors\cite{mario}.
Therefore, the different
dependences with energy of the these factors can
be checked against the experiment through comparison
of I(V) curves.

We apply our formalism to compute the I(V) curves
of the Au/Si(111) ($75 \AA$) and Au/Si(100) ($100 \AA$)
taken at low temperature (77 K)\cite{bell111,bell100}.
These are favourable cases to be compared with a
pure elastic theory, and in some way mark the
correctness of the name ballistic given to the
technique: from a direct comparison between
experiment and theory (e.g. see Fig. 4) we  
conclude that the ballistic assumption breaks down
at voltages higher than $1.3$ eV.
The theory has been calculated using k-space
distributions like the ones shown in Fig. 3,
with a very small $\eta$ (0.001 eV), and
assuming that the electrons are injected only
in the first pass.
The introduction of an electron-electron
inelastic interaction has the main effect
of taking out intensity due to the important
loss involved (typically half of the energy
is lost). In Fig. 4 we also show the effect of
introducing such a finite $\eta$ to match
the ballistic intensity to the experiment.
Our best fit has been obtained using the
energy dependence proposed by Bell\cite{bell111}
and for a mean free path of $155$ \AA\
at $1$ eV. While not a particular effort
to optimize these values has been made, 
both
$D=75$ \AA\ and $D=300$ \AA\ 
can be explained at the same time using
this approach. It is
worth commenting that the standard mean
free paths values derived from the
gas electron theory\cite{quinn} cannot describe correctly
the data because apparently underestimate
the electron-electron interaction resulting
in longer mean free paths.
Finally we should comment on the role of
the multiple reflections inside the metalic
layer. As mentioned above, the ballistic
result is obtained from direct injection,
but in a more appropriated description
multiple reflections must be taken into
account, as shown by Bell\cite{bell111}. 
Because mean free paths are longer
at lower energies their effect is more
important near the threshold, but on average
for a width like $D=75$ \AA\
three to four reflections are enough.
This brings about the model for reflections
at the surface and at the interface. We
have considered perfectly specular reflection
and completely diffuse reflection: differences
are not dramatic because of the focusing
effects introduced by the lattice, but
the diffuse reflection model produces
the best agreement with the experiments.

\begin{figure}
\epsfig{figure=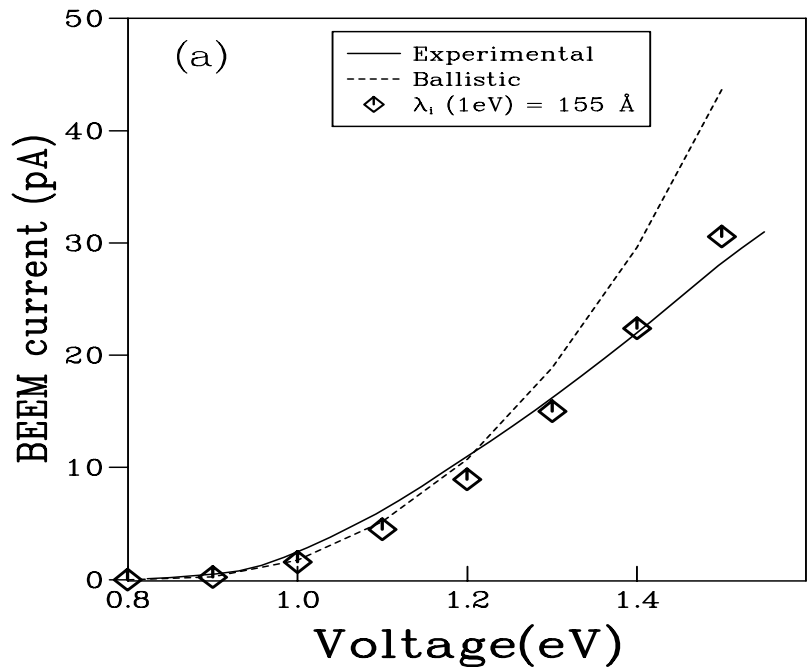,height=15cm,width=15cm}
\caption{
IV curve for Au on Si(111) under ballistic
conditions (dashed line), and including
inelastic effects (rhombes), compared
to experimental results (solid line)
as measured by
Bell in ref. [3] (D=75 A, T=77 K).
}
\end{figure}

\begin{figure}
\epsfig{figure=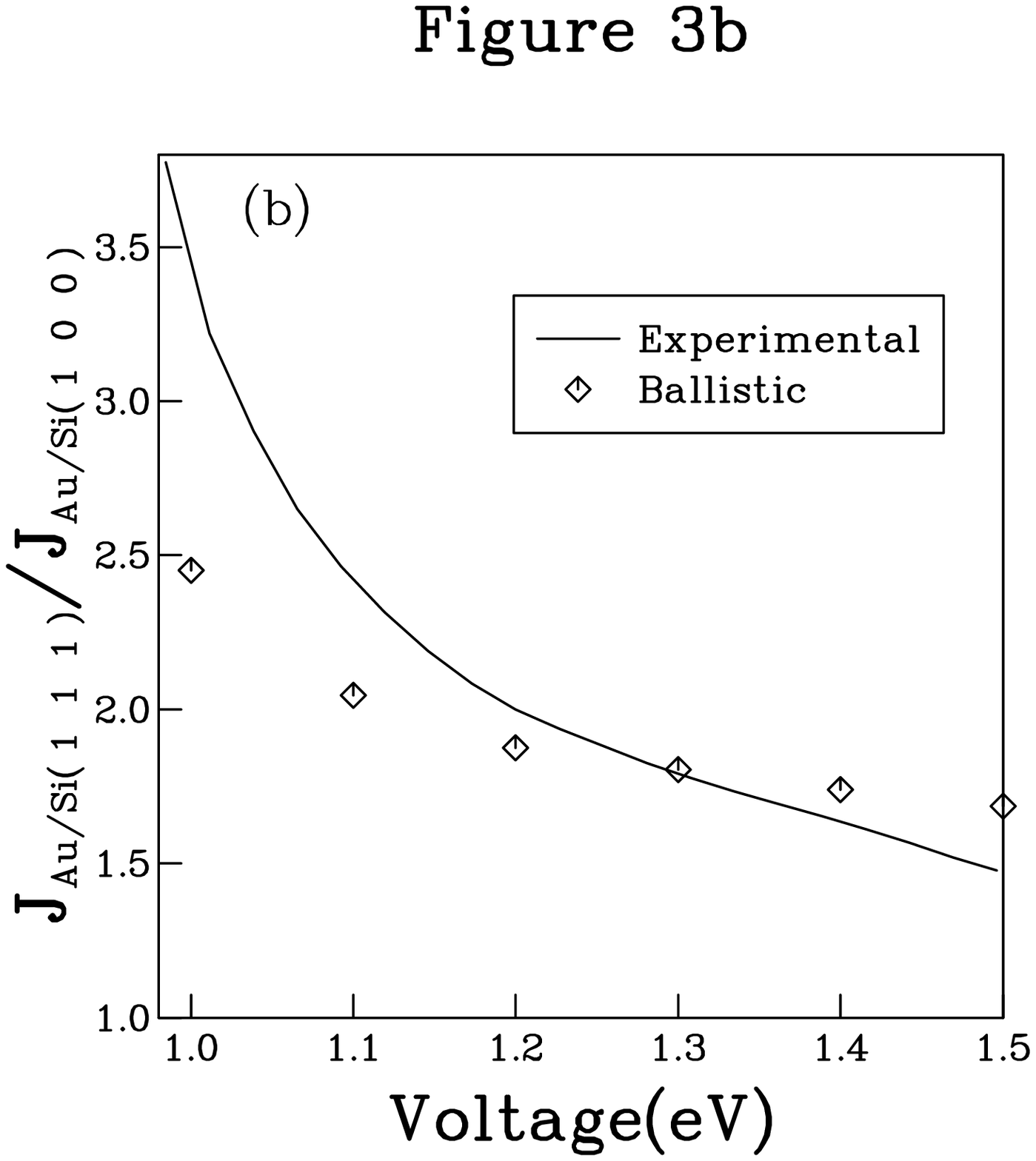,height=15cm,width=15cm}
\caption{
Comparison between BEEM currents
for Au on Si(111) (D=75 A) and Au on Si(100) (D=100 A).
Solid line: experimental data from
references [3] and [19].
Rhombes: J(100)/J(111)
computed from a ballistic theory.
}
\end{figure}

In the past, it has been difficult to understand
why Au/Si(111) and Au/Si(100) produced so similar
BEEM currents in spite of their different
projected band structure. This is specially
difficult without realizing
that the elastic interaction of electrons
with the lattice results in k-space current
distributions that accumulates around the two-dimensional
Brillouin zone edge. To test the similarity
between the two interfaces we compare in
Fig 5 our results for the two orientations
with the experiments. The ratio between
the intensity for the 111 orientation to
the intensity for the 100 orientation 
is plotted against the tip voltage. The same trend is
clearly observed in theory and experiment, 
and the main discrepancy (about $\frac{30}{100}$) 
is seen for low voltages, where the different
Schottky barrier for the two orientations 
produce a bigger uncertainty on the comparison.
This clearly shows how the different
projected band structure of the two orientations
for silicon yields similar BEEM-currents
without having to resort to non-conserving
parallel momentum due to scattering processes
of the electrons at the interface\cite{ludeke}.
 
\section{CONCLUSIONS}
We present an {\it ab initio}
theory based on Green's functions techniques
and a tight-binding 
hamiltonian to compute
the current distributions for a BEEM experiment in 
real and reciprocal space. Focusing of the
propagating electrons is determined by the
elastic interaction with the periodic lattice.
The resulting narrow lines and beams are of the
order of 2-3 atomic interspacing, explaining
the observed nanometric resolution of the technique.
In reciprocal space we observe a six-fold
distribution that determines the pure ballistic
current. Comparison of experimental results 
taken on thin layers and at low temperature
with the elastic theory ($\eta=0.001$ eV)
shows the influence of inelastic processes
in the high energy range ($\approx 1.5$ eV).
Finally, an inelastic electron-electron interaction
is taken into account resulting in a good
agreement between the theory and the
experimental data.

\section{ACKNOWLEDGMENTS}

We acknowledge financial support from the Spanish CICYT under
contracts number PB94-53 and PB92-0168C.
K.R. is grateful for financial support from
SFB292 (Germany).
We are grateful with Prof. P. Kocevar and
Dr. U. Hohenester for many interesting discussions
and their effort to build a {\it first-principles}
k-space Monte-Carlo approach to the BEEM problem,
and with Prof. K. Heinz for his continued
interest.

\end{document}